\documentclass[a4paper,11pt]{article}
\usepackage{pos}

\title{Optical spectroscopy of the VHE flat-spectrum radio quasar PKS~0903$-$57}

\author*[a]{E. Kasai}
\author[b]{P. Goldoni}
\author[c]{C. Boisson}
\author[d]{S. Pita}
\author[e]{W. Max-Moerbeck}
\author[f]{F. D'Ammando}
\author[e]{B. Rajput}
\author[a,g]{M. Backes}

\affiliation[a]{Department of Physics, Chemistry \& Material Science, University of Namibia, Private Bag 13301,\\
Windhoek, Namibia}

\affiliation[b]{Université Paris Cité, CNRS, CEA, Astroparticule et Cosmologie, F-75013 Paris, France}

\affiliation[c]{Laboratoire Univers et Théories, Observatoire de Paris, Université PSL, Université Paris Cité,\\ 
CNRS, F-92190 Meudon, France}

\affiliation[d]{Université Paris Cité, CEA, Astroparticule et Cosmologie, F-75013 Paris, France}

\affiliation[e]{Departamento de Astronomía, Universidad de Chile, Camino El Observatorio 1515, Las Condes, Santiago, Chile}

\affiliation[f]{INAF – Istituto di Radioastronomia, Via Gobetti 101, I-40129 Bologna, Italy}

\affiliation[g]{Centre for Space Research, North-West University, Private Bag X6001, Potchefstroom 2520,\\
South Africa}

\emailAdd{ekasai@unam.na}

\abstract{We report new optical spectroscopic results for the gamma-ray blazar PKS~0903$-$57. The optical properties of this source have historically been complicated to establish, due to a bright foreground star separated from the blazar by only 0.67$^{\prime\prime}$. During its strong high-energy (HE) and very high-energy (VHE) gamma-ray flaring period between 2020 and 2022, the Southern African Large Telescope and Very Large Telescope (VLT) took spectra of its optical counterpart. Photometric monitoring with the Rapid Eye Mount (REM) telescope complemented the observations. The observational results were inconclusive. A major improvement was achieved with a repeat of the VLT observations in 2024 using a 0.5$^{\prime\prime}$ slit under excellent seeing conditions. This led to the detection of five narrow emission features in the spectra, yielding a robust redshift measurement of $z$ = 0.2621 $\pm$ 0.0006. We also detected a broad symmetric H$\alpha$ emission component with a full width at half maximum of 4020 $\pm$ 30 km/s in the spectra. Non-thermal jet emission was found to dominate the continuum, supporting the classification of PKS~0903$-$57 as a flat-spectrum radio quasar. Another interesting observation was a significant velocity displacement of $\sim$1500~km/s between the broad H$\alpha$ line and its corresponding narrow line. Such a displacement may be associated with unusual dynamics in the broad-line region or accretion flow, a system containing two supermassive black holes, or a recoiling black hole produced following a supermassive black-hole merger. Building on these results, we carried out additional VLT, REM, and \emph{Swift} observations towards the end of 2024 and during 2025. We present in this paper the preliminary results from the analysis of these new data.}

\FullConference{African Astronomical Society Annual conference 2026 (AfAS2026)\\
22nd - 27th March 2026\\
Kasane, Botswana\\}

\begin{document}
\maketitle

\section{Introduction}
Blazars constitute one of the most extreme classes of active galactic nuclei (AGNs), displaying powerful emission that can extend from radio wavelengths to the HE (100~MeV~$\leq~E~\leq$~100~GeV) and VHE (100~GeV~$\leq E \leq$~10~TeV) ) gamma-ray domains. Their observed properties are generally attributed to relativistic plasma jets oriented at small angles to the observer's line of sight, resulting in strong relativistic beaming of the jet emission. Blazars are divided into two subclasses: BL Lacertae (BL Lac) objects and flat-spectrum radio quasars (FSRQs). In the optical regime, BL Lac objects typically exhibit weak or absent emission and absorption features, whereas FSRQs are distinguished by prominent emission lines superimposed on their continuum spectra.

PKS~0903$-$57 was originally identified as a radio source during the Parkes surveys conducted in the 1960s, within the declination range $-20^\circ$ to $+60^\circ$ [1]. Its gamma-ray emission was subsequently detected by the Large Area Telescope (LAT) aboard the \emph{Fermi Gamma-ray Space Telescope}, and the source was included in the first \emph{Fermi}-LAT source catalogue (1FGL) [2]. Since then, the source has been detected and included in every \emph{Fermi}-LAT catalogue.  In a recent paper by [3], PKS~0903$-$57 is identified as one of the brightest and hardest \emph{Fermi}-LAT FSRQs.

Determining the optical spectral properties of this source has been complicated by the presence of a nearby star at an angular separation of only 0.67$^{\prime\prime}$. This hindered efforts to establish the source's optical properties and reliably determine its redshift over the years. The close proximity resulted in the observed spectra containing a substantial flux contribution from the star, resulting in the masking or dilution of the intrinsic spectral signatures of the blazar. More recently, VLT spectroscopy obtained under subarcsecond seeing conditions provided an opportunity to substantially reduce the contribution from the star, allowing a more reliable investigation of the intrinsic optical spectrum of PKS~0903$-$57 [4]. The spectral investigations of the data enabled a robust redshift measurement of PKS~0903$-$57, and an identification of both narrow and broad emission-line features associated with it. These results and their astrophysical implications were reported and discussed in previous work [4,5]. The most peculiar feature of the system is the velocity offset ($\sim$ 1500~km/s) between the broad H$\alpha$ line and the narrow emission lines.

Building on such findings, a new multiwavelength observing campaign was undertaken during December 2024 and the first quarter of 2025, incorporating further observations with the VLT, REM telescope, and the \emph{Neil Gehrels Swift Observatory}. The main goals of these observations were to confirm the detection of the emission features, to check for variability in the broad emission line, in particular of its centroid, and to build a Spectral Energy Distribution (SED) of the source. In this paper, we extend the previous investigation of PKS~0903$-$57 and present preliminary results from the analysis of the new VLT, REM, and \emph{Swift} observations. 

\section{Observations and data reduction}
We performed three observations from December 2024 to March 2025 using the FOcal Reducer and low dispersion Spectrograph (FORS2) installed 
on the VLT. The observations were performed during exceptional seeing 
($\le 0.66^{\prime\prime}$) and using the position angle PA=108$^{\circ}$ as in our previous observations. The details of these observations are listed in Table~\ref{tabfors}. 

 \begin{table}[htp]
\caption{Parameters of the FORS2 observations of PKS~0903$-$57 presented in this contribution.}
\begin{center}
\begin{tabular}{cccccccc}
\hline
Instrument     & Range          & Resol. & Slit                     & Start Time                      & Exp   & Airmass & Seeing \\
        &   (\AA)          &    ($\lambda/\Delta\lambda$)       & ($^{\prime\prime}$) &  (UTC)                           & (s)     &              &($^{\prime\prime}$) \\
\hline \hline                     
FORS/600RI & 5200--8300  &  1000         &      0.5                 &  2024-12-29T06:41:35  & 2700  & 1.19     &   0.66 \\
FORS/600RI & 5200--8300  & 1000          &      0.5                 & 2025-01-23T06:03:47   &  2700 & 1.20     &    0.59 \\
FORS/1028z & 7800--9000  & 2500          &      0.5                 & 2025-03-28T01:52:41   & 2700  &  1.21    &    0.57 \\

\hline

\end{tabular}
\end{center}
\label{tabfors}
\end{table}%

The observations were designed to probe the variability (if any) of properties of the emission lines previously detected and were performed in service mode. In order to produce a more complete picture of the emission, we also organised a campaign of \emph{Swift} UV--X-ray and REM optical--NIR monitoring. The \emph{Swift} observations were triggered following the spectroscopic observations and were therefore conducted 2--3 days after the corresponding spectroscopic observations. REM observations are part of a long-term weekly monitoring program, which shows that PKS~0903$-$57 is highly variable (see Fig.~\ref{fig_lc_sed}, left panel). The contribution of the nearby star was subtracted from the UV, optical, and NIR photometric data. The data reduction, flux calibration, and telluric corrections of the FORS2 observations were performed as in our previous works (see, e.g., [6,7]). Given the contamination with the nearby star, we scaled the flux calibration of the spectrum using the near-simultaneous REM photometry corrected with the photometric template. 
 We performed data reduction of \emph{Swift} XRT and UVOT data using the HEASOFT software\footnote{https://swift.gsfc.nasa.gov/analysis/start/}. X-ray spectra were consistent with a power-law and energy fluxes were derived using a Galactic column density of $N_H$ = 2.6 $\times$ 10$^{21}$~cm$^{-2}$. We reduced all raw optical frames obtained with the REM telescope in the r filter following standard procedures. Instrumental magnitudes were obtained via aperture photometry and absolute calibration was performed by means of secondary standard stars in the field reported by the AAVSO Photometric All-Sky Survey (APASS) catalog\footnote{https://www.aavso.org/apass}.

\section{Data analysis and results}

\begin{figure}
     \centering
\includegraphics[width=7.3truecm,height=5.5truecm]{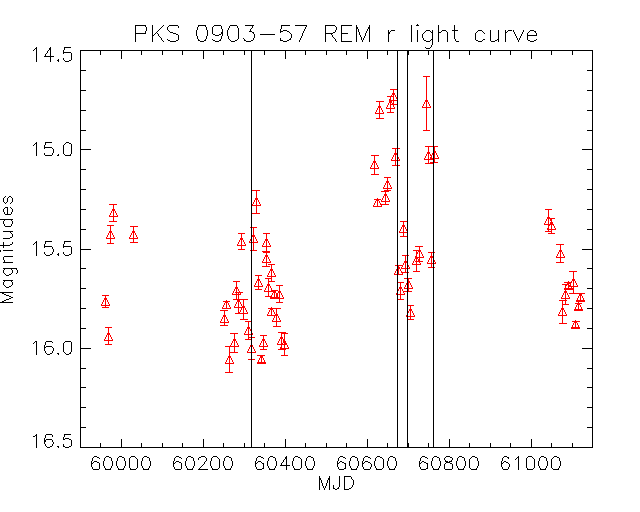} 
 \includegraphics[width=7.3truecm,height=5.5truecm]{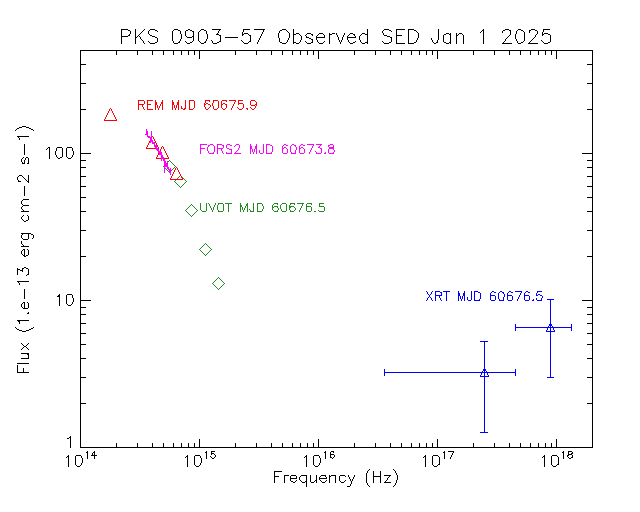}
 \caption{\emph{Left panel:} Long term PKS~0903$-$57 REM light curve (r-band) after star subtraction. The vertical lines indicate the times of the four FORS2 observations we performed. The first one reported in [4] and the three others reported in this contribution. \emph{Right panel:} Observed NIR-optical-X-ray SED of PKS~0903$-$57 at the beginning of January 2025. \emph{Swift} observations were performed on January~1, 2025, and REM observations on December~31, 2024. The brightest REM photometric point on the left corresponds to the H filter at frequency $\nu_H \sim$ 1.8 $\times$ 10$^{14}$~Hz and it seems to indicate a change in slope with a possible peak at that frequency or smaller. If so, it would suggest that the source is an ISP or an LSP (low-synchrotron-peaked)  blazar.}
     \label{fig_lc_sed}
\end{figure}

In the right panel of Figure~\ref{fig_lc_sed}, we present the observed spectral energy distribution (SED) of PKS~0903$-$57 around January 2025. 

The overall shape of the SED is similar to that observed during the 2020 outburst [8] with a peak in the NIR domain at $\nu F_\nu\sim 2\times 10^{-11}$~erg~cm$^{-2}$~s$^{-1}$, suggesting that the source, in this frequency range, has a similar level of activity. Note that in the \emph{Gaia} archive the source has $G$ = 17.9, roughly equivalent to $\nu F_\nu\sim10^{-12}$~erg~cm$^{-2}$~s$^{-1}$, therefore the source exhibits large long term variability. The position of the peak of the SED lies at frequencies smaller than 2 $\times$ 10$^{14}$ Hz, suggesting that PKS~0903$-$57 may be an intermediate-synchrotron-peaked (ISP) blazar.

\begin{figure*}
   \centering
 \includegraphics[width=7.3truecm,height=5.5truecm]{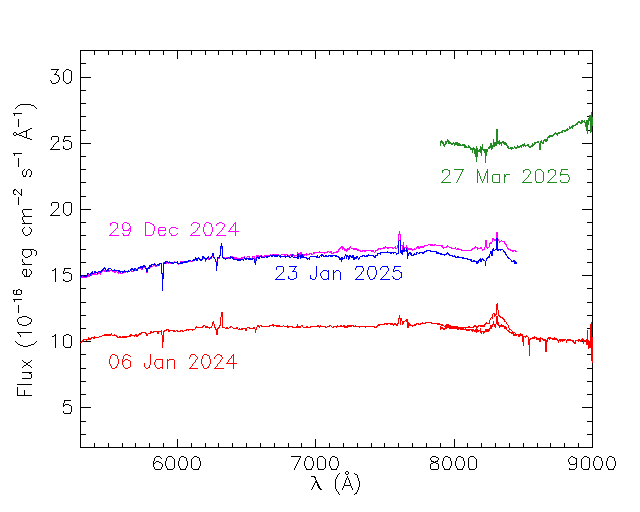} 
 \includegraphics[width=7.3truecm,height=5.5truecm]{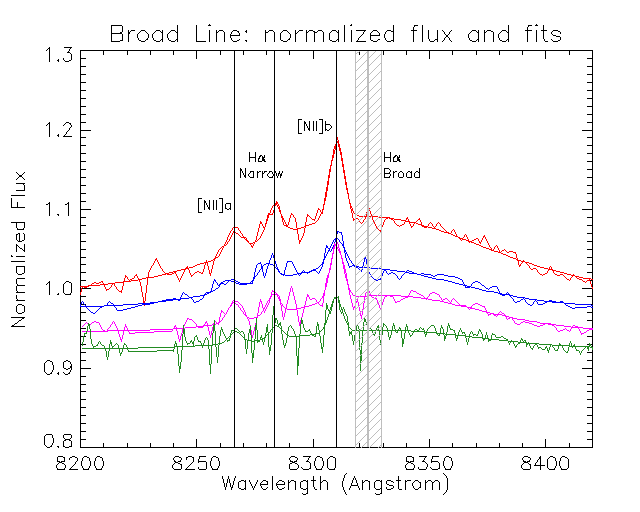} 

  \caption{\emph{Left panel:} Flux calibrated spectra of the four FORS2 VLT observations (including the already published one) with dates. Note that, in the three latest observations, the blazar was significantly brighter, especially in the latest. \emph{Right panel:} Normalized spectra of the four FORS2 observations in the region around the H$\alpha$-[NII] complex shifted along the vertical axis for better visibility. The positions of the narrow lines are marked with vertical lines. A grey shaded box centered on the January 2024 position of the broad line marks the range of positions consistent (within $\pm$ 200~km/s) with the broad lines in the observations reported here.}
 \label{fig_spectra_norm}
    \end{figure*}

The FORS2 spectra we obtained in this campaign display the same spectral features that were detected in our first observation in January 2024 [4]. However, as visible in Figure~\ref{fig_spectra_norm}, left panel, the blazar continuum was much brighter by a factor 1.5 to 2.5. The emission lines are therefore less prominent in these spectra. Indeed, while the narrow lines [OIII]b and [NII]b are clearly detected in all spectra, [OIII]a, [NII]a, and the narrow H$\alpha$ are only marginally detected.

  In our preliminary analysis the equivalent width of the broad H$\alpha$ line is always weaker than in the January 2024 observation and it decreases from 5.5 $\pm$ 0.2~\AA\ in the first spectrum to 2.4 $\pm$ 0.2~\AA\ in the third (see Table~\ref{tabew}). This is consistent with the overall brightening of the continuum emission as the main reason for this trend. The [OIII]b line, on the other hand, is weaker than in January 2024 but seems to stay constant or possibly grow in these observations.

  Finally, we investigated the velocity offset of the broad H$\alpha$ line. Due to the lower signal-to-noise of the line and the higher telluric contamination on the blue side of the line, the localization precision is lower than in the January 2024 observation. Performing gaussian fits of the feature, the errors on the positions of the centroids are of the order of 50--100~km/s (1$\sigma$). Preliminary analysis shows that the offsets are consistent within 2$\sigma$ with that previously reported, in the range 1,300-1,700~km/s. However systematic residuals are still present in the fits. This will be investigated in more detail to obtain better precision. 
  
\begin{table}[htp]
\caption{Equivalent width values of the broad H$\alpha$ and [OIII]b lines in our FORS2 spectra. While the broad H$\alpha$ detected flux is clearly decreasing, the [OIII]b detected flux after decreasing from the first observation, is constant or growing in the second and third spectrum (the fourth spectrum does not cover the [OIII] doublet).}
\begin{center}
\begin{tabular}{ccccc}
\hline\hline
Line      & Jan 2024       & Dec 2024       & Jan 2025         &    Mar 2025 \\
(1) & (2)& (3) & (4) & (5) \\ 
\hline
H$\alpha$   & 7.9 $\pm$ 0.2  & 5.5 $\pm$ 0.2 & 4.8 $\pm$ 0.2 & 2.4  $\pm$ 0.2 \\

[OIII]b & 1.4 $\pm$ 0.1 &  0.7 $\pm$ 0.1 &   0.9 $\pm$ 0.1  &  N/A \\  
\hline
\end{tabular}
\end{center}
\label{tabew}
\begin{flushleft}
{\bf Notes.} The columns are: (1) Emission line name, (2) to (5) date of observation (in the heading) and measured EW, where applicable.
\end{flushleft}
\end{table}%


\section{Discussion and conclusions}

  We report on optical spectroscopic (VLT/FORS2), UV-optical-NIR (\emph{Swift}/UVOT-REM) photometric and X-ray (\emph{Swift}/XRT) observations of the TeV FSRQ PKS~0903$-$57, mainly aimed at confirming the detection of the emission features and their properties. The photometric observations show that the source was about 1.5 -- 2.5 times brighter than in the January 2024 observation, and furthermore, shows strong short-term variability.  They also allowed for 
  the creation of an SED, which turned out to be 
  consistent with the one reported during the 2020 outburst. Interestingly, the optical flux level appears to be about 10 times higher than the one detected during \emph{Gaia} DR3 epochs (July 2014 -- May 2017), suggesting strong long-term variability.
  In the FORS2 spectra, we confirm the presence of broad and narrow spectral features.  The redshift of the source is consistent, albeit with a lower significance as a result of the stronger continuum. The broad H$\alpha$ line is also less intense, its position is broadly consistent (within 200~km/s) with the one measured previously. 
  
  The results we present confirm the picture of PKS~0903$-$57 as a highly variable FSRQ with an offset, symmetric broad H$\alpha$ line. No intrinsic variation of the line can be claimed in position, shape, or flux. This seems to disfavor the scenario of a peculiar accretion configuration where variations would be expected on the timescale of a few days [9], however, we cannot exclude this interpretation based on these data. The lack of variations on months and year timescales is well compatible with a supermassive black hole binary (SMBHB) in the process of merging, where the narrow lines are produced around one black hole and the broad line around another. In this case, no velocity variation is expected to be detectable on timescales of five years or less [10], for example, recently [11] reported the possible detection of orbital motion in the H$\beta$ line of the quasar J0950+5128 over 22~years of observations. Similarly, the scenario where the broad line is produced in a recoiling black hole formed after the coalescence of two black holes and which is being displaced or ejected from the center of the host galaxy, is compatible with our observations.
  
  Further observations may help decide which scenario is more appropriate. In particular, one may search for and measure additional broad lines such as MgII to see if their position is consistent with that of the broad H$\alpha$ or with that of the narrow lines. In the former case, the SMBHB scenarios would be strengthened. We remark, however, that in blazars the detection of the spectral features strongly depends on the brightness of the power-law continuum (see e.g.[12,13]). Therefore, a period of lower activity of the source would be ideal to investigate these issues.


\begin{thebibliography}{99}
\bibitem{}
Bolton, J. G., Gardner, F. F., \& Mackey, M. B., 1964, Australian Journal of Physics, 17, 340.
\bibitem{}
Abdo, A. A., Ackermann, M., Ajello, M., Allafort, A., Antolini, E., Atwood, W. B., ... \& Nakamori, T., 2010, The Astrophysical Journal Supplement Series, 188(2), 405.
\bibitem{}
Mohammed, P. N. N., Paliya, V. S., Ravikumar, C. D., 2026, JHEAP, 54, 100696.
\bibitem{}
Goldoni, P., Boisson, C., Pita, S., D’Ammando, F., Kasai, E., Max-Moerbeck, W., Backes, M., Cotter, G., 2024, A\&A, 691, L5.
\bibitem{}
Kasai, E., Goldoni, P., Pita, S., et al., Proceedings of the International Astronomical Union. 2021;17(S375):96-100.
\bibitem{}
Goldoni, P., Boisson, C., Pita, S., et al., 2021, A\& A, 650, A106
\bibitem{}
Kasai, E., Goldoni, P., Pita, S., et al., 2023, MNRAS, 518, 2675
\bibitem{}
Acharyya, A., Aharonian, F., Ayt Benkhali, F., et al., 2026, JHEAP, 53, 100599
\bibitem{}
Wang, J.-M., Du, P., Brotherton, M. S., et al., 2017, Nauture Astronomy, 1, 755
\bibitem{}
Kelley, L., Z., 2021, MNRAS, 500, 4065
\bibitem{}
Mohammed, N. N., Runnoe, J.C., Eracleous, M., et al., 2026, ApJ, 998, 286
\bibitem{}
D'Ammando, F, Goldoni, P., Max-Moerbeck, W., et al., 2024, A\&A, 683, A222
\bibitem{}
Rajput, B., Goldoni, P., Max-Moerbeck, W., et al., 2025, A\&A, 704, 190
\end{thebibliography}
\end{document}